\documentclass{elsart}
\usepackage{graphicx}
\usepackage{amssymb}
\newcommand{\beq}{\begin{eqnarray}}
\newcommand{\eeq}{\end{eqnarray}}

\begin{document}

\begin{frontmatter}

\title{A new method for studying the vibration of non-homogeneous membranes}
\author{Paolo Amore}
\ead{paolo.amore@gmail.com}
\address{Facultad de Ciencias, Universidad de Colima,\\
Bernal D\'{i}az del Castillo 340, Colima, Colima, Mexico}
\address{and \\ Physics Department, University of Texas at El Paso, \\
El Paso, Texas, USA}

\begin{abstract}
We present a method to solve the Helmholtz equation for a non-homogeneous membrane with Dirichlet boundary
conditions at the border of arbitrary two-dimensional domains. The method uses a collocation approach based 
on a set of localized functions, called "little sinc functions", which are used to discretize two-dimensional
regions. We have performed extensive numerical tests and we have compared the results obtained with the present 
method with the ones available from the literature. Our results show that the present method is very accurate and 
that its implementation for general problems is straightforward.
\end{abstract}


\end{frontmatter}


\section{Introduction}
\label{intro}

This paper focuses on the solution of the inhomogeneous Helmholtz equation 
\beq
- \Delta u(x,y) = E \rho(x,y) u(x,y)
\eeq
over an arbitrary two dimensional membrane $\mathcal{B}$, with Dirichlet 
boundary conditions at the border, $\partial \mathcal{B}$. $u(x,y)$ is the transverse displacement
and $E=\omega^2$, $\omega$ being the frequency of vibration of the membrane.

This problem has been considered in the past by several authors, using different techniques:
for example, Masad in \cite{Masad96} has studied the vibrations of a rectangular membrane 
with linearly varying density using a finite difference scheme and an approach based on perturbation
theory; the same problem was also considered by Laura and collaborators in \cite{Laura97}, using
an optimized Galerkin-Kantorovich approach and by Ho and Chen, \cite{Ho00}, who have used a hybrid method. 
Recently Reutskiy has put forward in \cite{Reut07} a new numerical
technique to study the vibrations of inhomogeneous membranes, the method of external and internal excitation.
Finally, Filipich and Rosales have studied the vibrations of membranes with a discontinous density profile.

In this paper we describe a different approach to this problem and  compare its performance with that
of the methods mentioned above. Our method is based on a collocation approach (see
for example \cite{Amore06}) which uses a particular set of functions,
the {\sl Little Sinc functions} (LSF), introduced  in \cite{Amore07a,Amore07b}, to obtain a
discretization of a finite region of the two-dimensional plane. These functions
have been used with success in the numerical solution of the Schr\"odinger equation
in one dimension, both for problems restricted to finite intervals and for problems
on the real line. In particular it has been observed that exponential convergence
to the exact solution can be reached when variational considerations are made 
(see \cite{Amore07a,Amore07b}). In ref.\cite{Amore07b}, the LSF were used to obtain a new 
representation for non--local operators on a grid and thus numerically solve the relativistic 
Schr\"odinger equation. An alternative representation for the quantum mechanical path integral was also given
in terms of the LSF.

Although Ref.\cite{Amore07a} contains a detailed discussion of the LSF, I will briefly
review here the main properties, which will be useful in the paper. Throughout the paper I will
follow the notation of \cite{Amore07a}.

A Little Sinc Function is obtained as an approximate representation of the Dirac delta function
in terms of the wave functions of a particle in a box (being $2L$ the size of the box). Straightforward
algebra leads to the expression
\beq
s_k(h,N,x) \equiv  \frac{1}{2 N} \ \left\{ \frac{\sin \left( (2N+1) \ \chi_-(x)\right)}{\sin \chi_-(x)}
-\frac{\cos\left((2N+1)  \chi_+(x)\right)}{\cos \chi_+(x)} \right\} \ .
\label{sincls}
\eeq
where $\chi_\pm(x) \equiv \frac{\pi}{2 N h} (x \pm k h)$. The index $k$ takes the integer values
between $-N/2+1$ and $N/2-1$ ($N$ being an even integer). The LSF corresponding to a specific 
value of $k$ is peaked at $x_k = 2 L k/N = k h$, $h$ being the grid spacing and $2L$ the total extension
of the interval where the function is defined. By direct inspection of eq.~(\ref{sincls}) it is found that
$s_k(h,N,x_j) = \delta_{kj}$, showing that the LSF takes its maximum value at the $k^{th}$ grid 
point and vanishes on the remaining points of the grid.

It can be easily proved that the different LSF corresponding to the same set are orthogonal~\cite{Amore07a}:
\beq
\int_{-L}^L s_k(h,N,x) s_j(h,N,x) dx = h \ \delta_{kj} 
\eeq
and that a function defined on $x\in (-L,L)$ may be approximated as 
\beq
f(x) \approx \sum_{k=-N/2+1}^{N/2-1} f(x_k) \ s_k(h, N,x) \ . \label{f(x)_LSF}
\eeq

This formula can be applied to obtain a representation of the derivative of a LSF in terms of the set of LSF
as:
\beq
\frac{d s_k(h,N,x)}{dx} \approx \sum_j \left. \frac{d s_k(h,N,x)}{dx}\right|_{x=x_j} \ s_j(h,N,x) 
\equiv \sum_j c_{kj}^{(1)} \ s_j(h,N,x) \nonumber \\
\frac{d^2 s_k(h,N,x)}{dx^2} \approx \sum_j  \left. \frac{d^2 s_k(h,N,x)}{dx^2}\right|_{x=x_j} \ s_j(h,N,x) \equiv 
\sum_j c_{kj}^{(2)} \ s_j(h,N,x)  \ , \nonumber \\
\label{der2}
\eeq
where the expressions for the coefficients $c_{kj}^{(r)}$ can be found in \cite{Amore07a}. Although eq.~(\ref{f(x)_LSF})
is approximate and the LSF strictly speaking do not form a basis, the error made with this approximation decreases with $N$ and tends to
zero as $N$ tends to infinity, as shown in \cite{Amore07a}. For this reason, the effect of this approximation is 
essentially to replace the continuum of a interval of size $2L$ on the real line with a discrete set of $N-1$ points, $x_k$,
uniformly spaced on this interval. 

Clearly these relations are easily generalized to functions of two or more variables. Since the focus of this paper is
on two dimensional membranes, we will briefly discuss how the LSF are used to discretize a region of the plane; the
extension to higher dimensional spaces is straightforward.  
A function of two variables can be approximated in terms of $(N_x-1) \times (N_y-1)$ functions, corresponding to  
the direct product of the $N_x-1$ and $N_y-1$ LSF in the $x$ and $y$ axis: each term in this set corresponds to a specific 
point on a rectangular grid with spacings $h_x$ and $h_y$ (in this paper we use a square grid with $N_x=N_y=N$ and $L_x = L_y=L$).

Since $(k,k')$ identifies a unique point on the grid, one can select this point using a single index
\beq
K &\equiv& k'+\frac{N}{2} + (N-1) \left(k+\frac{N}{2}-1\right) \label{KK} 
\eeq
which can take the values $1 \leq K \leq (N-1)^2$. This relation can be inverted to give
\beq
k &=& 1 - N/2 + \left[\frac{K}{N-1+\varepsilon}\right] \\ 
k' &=& K - N/2 - (N-1) \ \left[ \frac{K}{N-1+\varepsilon}\right] \ ,
\eeq
where $\left[ a \right]$ is the integer part of a real number $a$ and $\varepsilon \rightarrow 0$.

To illustrate the collocation procedure we can consider the Schr\"odinger equation in two dimensions:
\beq
\hat{H} \psi_n(x,y) \equiv \left[- \Delta + V(x,y) \right] \psi_n(x,y) = E_n \psi_n(x,y) 
\label{schrodinger}
\eeq
using the convention of assuming a particle of mass $m=1/2$ and setting $\hbar = 1$. The Helmholtz
equation, which describes the vibration of a membrane, is a special case of (\ref{schrodinger}), corresponding
to having $V(x,y) = 0$ inside the region $\mathcal{B}$ where the membrane lies and $V(x,y)=\infty$ on the border $\partial \mathcal{B}$
and outside the membrane. 

The discretization of eq.~(\ref{schrodinger}) proceeds in a simple way using the properties 
discussed in eqs.~(\ref{f(x)_LSF}) and (\ref{der2}):
\beq
H_{kk',jj'} =  -\left[ c^{(2)}_{kj} \delta_{k'j'} + \delta_{kj} c^{(2)}_{k'j'} \right]+ \delta_{kj} \delta_{k'j'} V(x_k,y_{k'}) 
\label{Hamiltonian}
\eeq
where $(k,j,k',j') = - N/2+1, \dots, N/2-1$. Notice that the potential part of the Hamiltonian is obtained 
by simply "collocating" the potential $V(x,y)$ on the grid, an operation with a limited computational
price. The result shown in (\ref{Hamiltonian}) corresponds to the matrix element of the Hamiltonian operator
$\hat{H}$ between two grid points, $(k,k')$ and $(j,j')$, which can be selected using two integer values $K$ and $J$, 
as shown in (\ref{KK}). 

Following this procedure the solution of the Schr\"odinger (Helmholtz) equation on the uniform grid generated by the LSF
corresponds to the diagonalization of a $(N-1)^2 \times (N-1)^2$ square matrix, whose elements are given by eq.~(\ref{Hamiltonian}).

\section{Applications}
\label{application}

In this Section we apply our method to study the vibration of different non-homogeneous membranes. Our examples are a
rectangular membrane with a linear and oscillatory  density, a rectangular membrane with a piecewise constant density, a
circular membrane with density $\rho(x,y) = 1 +\sqrt{x^2+y^2}$ and a square membrane with a variable density which fluctuates randomly
around a constant value. All the numerical calculations have been performed using Mathematica 6.

\subsection{A rectangular membrane with linearly varying density}
\label{rect1}

Our first example is taken from \cite{Masad96} and later studied by different authors \cite{Laura97,Ho00,Reut07}; 
these authors have considered the Helmholtz equation over a rectangle of sides
$a$ and $b$, and with a density
\beq
\rho(x,y) = 1+\alpha \left(\frac{x}{a} +\frac{1}{2}\right) \ .
\eeq

Notice that the factor $1/2$ appearing in the expression above derives from our convention of centering the rectangle in the origin, whereas
the authors of \cite{Masad96,Laura97} consider the regions $x \in (0,a)$ and $y \in (0,b)$. 

The Helmholtz equation for an inhomogeneous membrane is 
\beq
-\Delta u(x,y) = \omega^2  \rho(x,y) u(x,y)
\label{helm1}
\eeq
where $u(x,y)$ is the transverse displacement and $\omega$ is the frequency of vibration.

As explained in \cite{Amore08}, the collocation of the inhomogeneous Helmholtz equation is straightforward, 
and in fact it does not require the calculation of any integral. The basic step is to rewrite eq.~(\ref{helm1}) 
into the equivalent form
\beq
- \frac{1}{\rho(x,y)} \ \Delta u(x,y) = \omega^2 u(x,y) \ .
\label{helm2}
\eeq

The operator $\hat{O} \equiv  -\frac{1}{\rho(x,y)} \ \Delta$ is easily collocated on the uniform grid generated by the LSF, and a 
matrix representation is obtained. To see how this is achieved we can limit ourselves to a
one dimensional operator and make it act over a single LSF:
\beq
-\frac{1}{\rho(x)} \frac{d^2}{dx^2} s_k(h,N,x) &=& - \sum_{jl} \frac{1}{\rho(x_j)} c^{(2)}_{kl} s_j(h,N,x) s_l(h,N,x) \nonumber \\
&\approx& - \sum_{j} \frac{1}{\rho(x_j)} c^{(2)}_{kj} s_j(h,N,x) \ .
\eeq

The matrix representation of this operator over the grid can now be read explicitly from the expression above. It is important to
notice that in general the matrix representation of $\hat{O}$ will not be symmetric, unless the membrane is homogeneous. 
From a computational point of view the diagonalization of symmetric matrices is typically faster than for non-symmetric matrices 
of equal dimension.

\begin{table}
\begin{center}
\begin{tabular}{|c|cc|cc|}
\hline
$n$ &  $N=10$ & $N=12$ & $N=10$ & $N=12$ \\
\hline
1&  4.335384404 & 4.335384227 & 3.610497303 & 3.610490268 \\
2&  6.853837244 & 6.853836545 & 5.670792660 & 5.670765953 \\
3&  6.856020159 & 6.856019330 & 5.755204660 & 5.755169941 \\
4&  8.672635329 & 8.672633652 & 7.290804774 & 7.290733907 \\
5&  9.690424142 & 9.690422167 & 7.942675413 & 7.942603436 \\
6&  9.696016734 & 9.696013642 & 8.146392098 & 8.146261520 \\
7&  11.05545104 & 11.05544646 & 9.299769374 & 9.299574120 \\
8&  11.05628129 & 11.05627781 & 9.305141208 & 9.304993459 \\
9&  12.63048727 & 12.63048290 & 10.24733083 & 10.24717742 \\
10& 12.64211438 & 12.64210427 & 10.62452059 & 10.62409026 \\
\hline
\end{tabular}
\end{center}
\bigskip
\caption{Results for the first $10$ frequencies for $b/a=1$ and $\alpha =0.1$ (second and third columns) and to $\alpha =1$ (fourth and fifth columns).}
\label{table1}
\end{table}

In Table \ref{table1} we display the first $10$ frequencies of the square membrane ($b/a=1$) for $\alpha=0.1$ (second and third columns) 
and for $\alpha =1$ (fourth and fifth columns). The number of collocation points is determined by the parameter $N$ which is fixed to 
$10$ (second and fourth columns) and to $12$ (third and fifth columns). The comparison of the results corresponding to different $N$ gives
us an information over the precision of the results: looking at the Table we see that typically the results agree at least in the first $5$
digits, although we are working with a rather sparse grid. Also it should be remarked that the method is providing a whole set of 
eigenvalues and eigenvectors, $(N-1)^2$ to be exact, whereas in other approaches each mode is studied separately.

To allow a comparison with the results of ref.~\cite{Masad96,Laura97,Ho00,Reut07} a calculation of the fundamental frequency of the rectangular
membrane for different sizes of the membrane and different density profile is reported in Table \ref{table2}. The numerical results have been
obtained working with $N=12$. These results can be compared with those of Table 1 and 2 of ref.~\cite{Masad96}, of Table 1 of ref.~\cite{Laura97}, of
Table 1 of ref.~\cite{Ho00} and of Table 10 of ref.~\cite{Reut07} (in the last two references only the case $\alpha=0.1$ is studied).
Comparing our results with those of Masad we have been able to confirm the observation of Laura et al., that the frequencies calculated by Masad
for $\alpha=1$ and $b/a=0.6,0.4,0.2$ are incorrect. Actually, the results reported by Masad in these cases are just the 
second frequency (for $b/a=0.6,0.4$) and the fourth frequency (for $b/a=0.2$) of the corresponding rectangular membrane, which the author 
failed to identify as such. This observation illustrates the advantage of working with a method which provides a tower of frequencies
at the same time.

\begin{figure}[t]
\begin{center}
\includegraphics[width=8cm]{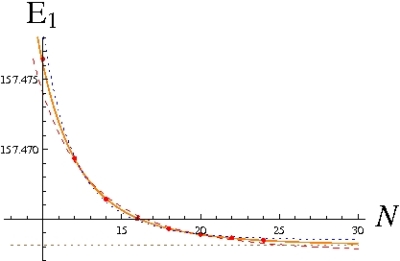}
\caption{Square of the fundamental frequency of the rectangular membrane for $b/a=0.2$ and $\alpha=1$ as a function of $N$. The dashed, solid and
dotted lines correspond to fitting the numerical points with $f_r(N) = c_{0} +c_{1}/N^r$ with $r=3,4,5$ respectively. The horizontal line
is the limit value of $f_4(N)$ for $N \rightarrow \infty$, corresponding to $E_1 = \omega_1^2 = 157.4631229$.}
\label{fig_1}
\end{center}
\end{figure}

Another great advantage of our method is the great rate of convergence which is typically observed as the number of grid points is increased.
In the cases studied in ref.~\cite{Amore08} where the boundary conditions are enforced exactly (a circle and a circular waveguide) we observed
that the leading non-constant behaviour of the eigenvalues for $N \gg 1$ was $1/N^4$.  This analysis in now repeated here for the 
case of $\alpha=1$ and $b/a=0.2$: the points in Fig.\ref{fig_1} correspond to the square of the fundamental frequency calculated using
different grid sizes. The curves which decay with $N$ correspond to fitting the numerical points with $f_r(N) = c_0 +c_1/N^r$ with $r=3,4,5$ 
respectively. The horizontal line is the limit value of $f_4(N)$ for $N \rightarrow \infty$, which corresponds to $\omega =12.548431091$ (notice
that the result obtained for $N=12$ agrees in its first five digit with this result, whereas the analogous result of ref.~\cite{Laura97} agrees
only in three digits). The reader will certainly notice that $f_4(N)$ fits excellently the sets, thus confirming the observations made in ref.~\cite{Amore08}. 
A further important observation concerns the monotonic behaviour of the points in the figure: as observed already in ref.~\cite{Amore08} this method
typically provides monotonic sequences of approximations which approach the exact value from above.

\begin{figure}[t]
\begin{center}
\includegraphics[width=8cm]{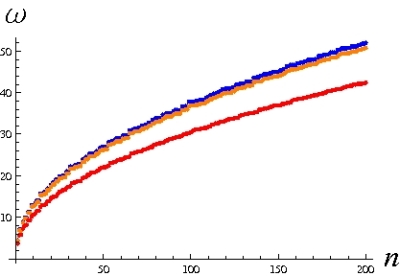}
\caption{First $200$ frequencies of a square membrane with density $\rho(x) = 1+ \alpha  (x+1/2)$  for 
$\alpha = 0, 0.1,1$ (from top to bottom) obtained using $N=26$.}
\label{fig_2}
\end{center}
\end{figure}

In Fig.~\ref{fig_2} we have plotted the first $200$ frequencies of a square membrane with $\alpha=0,0.1,1$ (going from top to bottom), using
a grid with $N=26$.

\subsection{A rectangular membrane with oscillating density}
\label{rect2}

As a second example, we consider a rectangular membrane with density $\rho(x) = 1+ 0.1 \sin\pi (x+1/2)$. This problem has been studied in 
ref.~\cite{Ho00,Reut07}. In Table \ref{table3} we compare our results (LSF) with the results of ref.~\cite{Ho00}, for different
sizes of the membrane. Our results, obtained with a grid corresponding to $N=20$, agree with those of ref.~\cite{Ho00}. 
In Table \ref{table4} we compare the results for the first $10$ frequency of the square membrane with this density with those reported in
ref.~\cite{Reut07}. Although our first $7$ results agree with those of ref.~\cite{Reut07}, we have noticed that the remaining results do not agree. 
By looking at the table the reader will notice that the disagreement is caused by the fact that ref.~\cite{Reut07} has missed a 
frequency (such an error cannot take place with our method, since the diagonalization of the Hamiltonian matrix automatically provides the
lowest part of the spectrum).

\begin{table}
\begin{center}
\begin{tabular}{|c|cc|}
\hline
$b/a$ &  $\alpha=0.1$ & $\alpha=1$ \\
\hline
1&  4.335384227 & 3.610490268 \\
0.8&  4.907186348 & 4.08151588  \\
0.6&  5.957896353 & 4.942230449 \\
0.4&  8.252203964 & 6.797302677 \\
0.2&  15.61334941 & 12.54867967 \\
\hline
\end{tabular}
\end{center}
\bigskip
\caption{Results for the fundamental frequency of a rectangular membrane with density $\rho(x) = 1+ \alpha  (x+1/2)$ 
for $\alpha =0.1$ and $\alpha =1$, using $N=12$.}
\label{table2}
\end{table}

\begin{table}
\begin{center}
\begin{tabular}{|c|cc|}
\hline
$b/a$ &  LSF & Ref.~\cite{Ho00} \\
\hline
1  &  4.265402726 & 4.26541\\
0.8&  4.828066678 & 4.82806\\
0.6&  5.862077020 & 5.86207\\
0.4&  8.120442809 & 8.12044\\
0.2&  15.37382214 & 15.37381\\
\hline
\end{tabular}
\end{center}
\bigskip
\caption{Results for the fundamental frequency of the rectangular membrane with density $\rho(x) = 1+ \alpha \sin\pi (x+1/2)$ 
for $\alpha =0.1$, using the LSF with $N=20$ (second column). The third column are the results of Ref.\cite{Ho00}.}
\label{table3}
\end{table}

\begin{table}
\begin{center}
\begin{tabular}{|c|cc|}
\hline
$n$ &  LSF & Ref.~\cite{Reut07} \\
\hline
1& 4.265402726& 4.265404 \\
2& 6.743887484& 6.743888 \\
3& 6.797319723& 6.797326 \\
4& 8.597648785& 8.597662 \\
5& 9.536589305& 9.536574 \\
6& 9.624841722& 9.624849 \\
7& 10.95914134& 10.95915 \\
8& 10.97412691& 12.43285 \\
9 & 12.43293737& 12.55436 \\
10& 12.55438511& 12.91349 \\
11& 12.91343471&  $-$ \\
\hline
\end{tabular}
\end{center}
\bigskip
\caption{First $11$ frequencies of a square membrane with density $\rho(x) = 1+ 0.1 \sin\pi (x+1/2)$ 
using the LSF with $N=20$ (second column). The third column are the results of Ref.\cite{Reut07}.}
\label{table4}
\end{table}

\begin{figure}[t]
\begin{center}
\includegraphics[width=8cm]{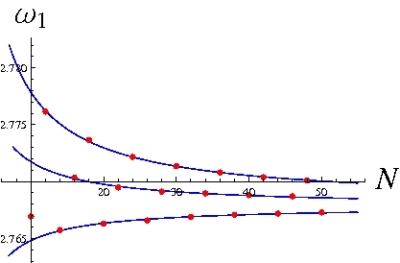}
\caption{Fundamental frequency of the rectangular membrane with discontinous density. The solid lines are the fits $f(N) = c_0+c_1/N$
over the three different monotonous sequences of numerical approximations.}
\label{fig_3}
\end{center}
\end{figure}

\subsection{A rectangular membrane with discontinous density}
\label{rect3}

Our next example is taken from ref.~\cite{Kang02,Filip07}: it is a rectangular membrane of sizes $a=1$ and $b=1.8$. The membrane is divided
into two regions by the line~\footnote{We use the convention of centering the membrane in the origin of the coordinate axes.}
\beq
y+b/2 = 0.3 (x+a/2)+0.7  \ .
\eeq 
The upper region has a density which is twice as big as the density of the lower region. The collocation procedure is the same as in the 
previous example. In Fig.~\ref{fig_3} we have plotted the fundamental frequency of the membrane calculated with different grid sizes, i.e.
different $N$. In this case we clearly observe an oscillation of the numerical value: as discussed in  ref.~\cite{Amore08} this behaviour 
is typical when the grid does not cross the boundary. Although in the present case the Dirichlet boundary conditions are enforced exactly
on the border of the rectangle, the region of discontinuity is not sampled optimally by the grid, which causes the oscillation.
Nonetheless, we may observe that the set of numerical values can be divided into three distinct and equally spaced  sets, each of 
which can be fitted quite well with a behaviour $f(N) = c_0+c_1/N$ (the solid curves in the plot).
Going from top to bottom, the curves correspond to the fits:
\beq
f(N) &=& 2.76810+ 0.0975751/N \\
f(N) &=& 2.76778+ 0.0395891/N \\
f(N) &=& 2.76785- 0.0304169/N \ .
\eeq

\begin{table}
\begin{center}
\begin{tabular}{|c|ccc|}
\hline
$n$ &  $N=34$ & $N=40$ & $N=46$  \\
\hline
1   &   2.768965729   &   2.768808666   &   2.768693679\\
2   &   3.967450487   &   3.967209040   &   3.966996848\\
3   &   4.845749110   &   4.845534206   &   4.845424548\\
4   &   5.006258593   &   5.006199469   &   5.006157369\\
5   &   5.844875008   &   5.844420528   &   5.844031012\\
6   &   6.308222252   &   6.307796274   &   6.307464549\\
7   &   6.926024314   &   6.925679216   &   6.925339604\\
8   &   7.006779901   &   7.006553463   &   7.006421938\\
9   &   7.632533775   &   7.632317904   &   7.632118114\\
10   &  7.711258827   &   7.711116778   &   7.711041692\\
\hline
\end{tabular}
\end{center}
\bigskip
\caption{First $10$ frequencies of a rectangular membrane with discontinous density for different grid sizes.}
\label{table5}
\end{table}

\subsection{A circular membrane with  density $\rho(x,y) = 1 +\sqrt{x^2+y^2}$}
\label{circ}

\begin{figure}[t]
\begin{center}
\includegraphics[width=8cm]{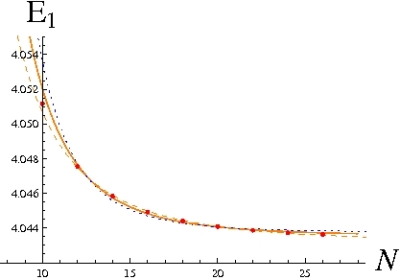}
\caption{Square of the fundamental frequency of the circular membrane with density $\rho(x,y) = 1 +\sqrt{x^2+y^2}$ as a function of $N$. 
The dashed, solid and dotted lines correspond to fitting the numerical points with $f_r(N) = c_{0} +c_{1}/N^r$ with $r=3,4,5$ respectively.}
\label{fig_5}
\end{center}
\end{figure}

Another interesting example is taken from ref. \cite{Reut07}, where an inhomogeneous circular membrane with  density $\rho(x,y) = 1 +\sqrt{x^2+y^2}$
is considered. Table 8 of ref.\cite{Reut07} contains the first $5$ eigenvalues. We have applied our method to this problem, using grids
of different sizes, with $N$ going from $10$ to $30$. Studying the $N$ dependence of the eigenvalues, we have seen that these decrease 
monotonically and that the leading non-constant dependence on $N$ for $N \rightarrow \infty$ is $N^{-3}$ (see Fig.\ref{fig_5}). 
In Table \ref{table6} we report the first $10$ frequencies of a
circular membrane with density  $\rho(x,y) = 1 +\sqrt{x^2+y^2}$ for different grid sizes ($N=26,28,30$). Notice that the results
already agree in their first $4$ digits. A more precise result is then obtained by performing an extrapolation of the numerical results
for grids going from $N=12$ to $N=30$. Notice the good agreement with the results of \cite{Reut07} (although we believe that our results are
more precise), and that, as expected, some frequencies are degenerate (the degeneracy of the frequencies is not discussed in  \cite{Reut07}).

\begin{figure}[t]
\begin{center}
\includegraphics[width=8cm]{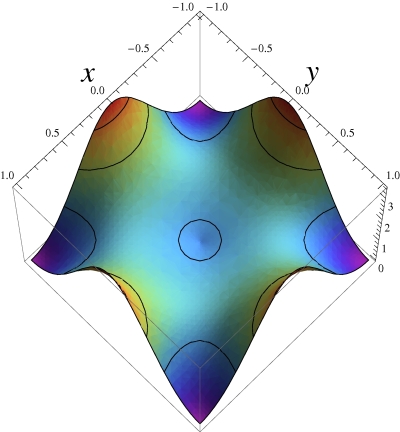}
\caption{Density of the square inhomogeneous membrane obtained by conformal mapping of the circular homogeneous membrane 
with density $\rho(x,y) = 1 +\sqrt{x^2+y^2}$.  The solid lines mark the level $1$,$2$ and $3$. }
\label{fig_4}
\end{center}
\end{figure}

\begin{table}
\begin{center}
\begin{tabular}{|c|ccccc|}
\hline
$n$ &  $N=26$ & $N=28$ & $N=30$ & extrapolated & Ref.\cite{Reut07}  \\
\hline
1   &   2.010879124   &   2.010861686   &   2.010848852   &   2.010797145 & 2.00987\\
2   &   3.067875783   &   3.067856983   &   3.067843839   &   3.067802071 & 3.06760\\
3   &   3.067875783   &   3.067856983   &   3.067843839   &   3.067802071 &  - \\
4   &   4.022444836   &   4.022423309   &   4.022408260   &   4.022360463  & 4.02232\\
5   &   4.022616933   &   4.022550844   &   4.022504777   &   4.022359634 & - \\
6   &   4.555479648   &   4.555348923   &   4.555254320   &   4.554900658  & 4.55457\\
7   &   4.928682458   &   4.928599914   &   4.928542452   &   4.928361919 & 4.92830\\
8   &   4.928682458   &   4.928599914   &   4.928542452   &   4.928361919 & - \\
9   &   5.699612500   &   5.699483990   &   5.699394977   &   5.699117837 & - \\
10   &  5.699612500   &   5.699483990   &   5.699394977   &   5.699117837 & - \\
\hline
\end{tabular}
\end{center}
\bigskip
\caption{First $10$ frequencies of a circular membrane with density $\rho(x,y) = 1 +\sqrt{x^2+y^2}$ for different grid sizes. The 
fourth column reports the frequencies obtained by extrapolating the numerical results for grids going from $N=12$ to $N=30$. 
The last column reports the results of ref.~\cite{Reut07}.}
\label{table6}
\end{table}

\subsection{A square membrane with a random density}
\label{random}

Our last example is a square membrane with a variable density which fluctuates randomly around the value $\rho_0=1$. We have
generated random values for the density on a uniform square grid with $81$ points (corresponding to $N_0=10$ in our notation); 
at these points the value of the density has been chosen according to the formula:
\beq
\rho(x_k,y_j) = \rho_0 + \delta \rho \ q_{kj} 
\eeq
where $\rho_0=1$ and $\delta\rho=1/2$. $q_{kj}$ is a random number distributed uniformly between $-1/2$ and $1/2$. The density 
over all the square has then been obtained interpolating with the LSF:
\beq
\rho(x,y) = \mathcal{C} \sum_{kj} \rho(x_k,y_j) \ s_k(h,N_0,x) s_j(h,N_0,y) \ ,
\eeq
where $N_0=10$, as previously mentioned. $\mathcal{C}$ is a normalization constant which constrains the total mass of the membrane
to be equal to the mass carried by the homogeneous membrane with density $\rho_0$.

\begin{figure}[t]
\begin{center}
\includegraphics[width=6.5cm]{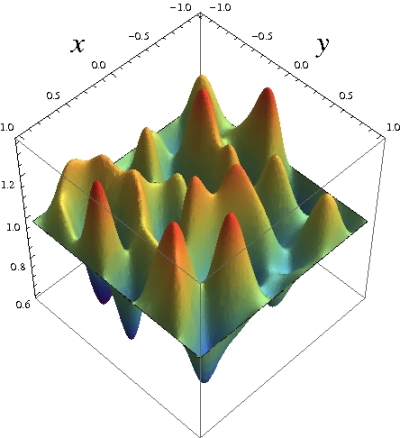}
\includegraphics[width=6.5cm]{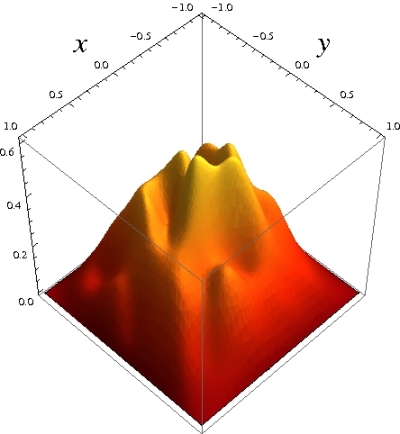}
\caption{Left Panel:Density of the square membrane with random density (Set 1). Right Panel:
Fundamental mode of the square membrane with random density (using Set 1). We have used the LSF on a grid with $N=40$. }
\label{fig_6}
\end{center}
\end{figure}

\begin{figure}[t]
\begin{center}
\includegraphics[width=10cm]{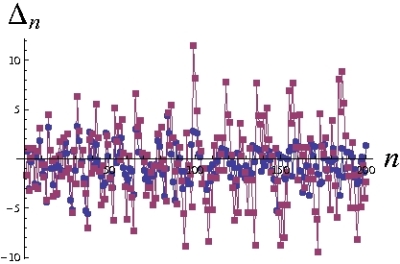}
\caption{$\Delta_n= E_n - E_n^{Weyl}$ for the homogeneous square membrane (squares) and the square membrane with randomly 
oscillating density (circles).}
\label{fig_7}
\end{center}
\end{figure}

In the left panel of Fig.\ref{fig_6} we plot the density of the membrane, while in the right plot we plot the fundamental mode.
We have performed our calculation using a grid with $N=40$ (i.e. with a total of $1521$ modes): upon diagonalization 
of the matrix obtained using the collocation procedure we have obtained numerical estimates for the lowest part of the spectrum
of the random membrane. These results may be compared with those of the homogeneous square membrane and with the asymptotic 
behavior predicted by Weyl's law \cite{Kut84}:
\beq
E_n^{Weyl} = \frac{4 \pi n}{A} + \frac{L}{A} \sqrt{\frac{4 \pi n}{A}} \ ,
\label{weyl}
\eeq
$A$ being the area and $L$ the perimeter of the membrane. In Fig.~\ref{fig_7} we plot the quantity 
$\Delta_n= E_n - E_n^{Weyl}$ for the homogeneous square membrane (squares) and the square membrane with randomly oscillating density (circles).
We have limited the plot to the first $200$ modes.
Notice that in both cases $\Delta$ oscillates around $0$, although the oscillation are smaller for the random membrane.

We have also fitted the first $400$ modes with the functional form given by Weyl's law obtaining
\beq
E_n \approx 3.14546 n+3.45835 \sqrt{n}
\eeq
for the homogeneous membrane and
\beq
E_n \approx 3.14445 n+3.44674 \sqrt{n}
\eeq
for the random membrane. These values should be compared with the one given by eq.~(\ref{weyl}):
\beq
E_n \approx 3.14159 n+3.54491 \sqrt{n} \ .
\eeq

\section{Conclusions}

In this paper we have introduced a new method to solve the Helmholtz equation for non-homogeneous membranes. This method
uses the Little Sinc Functions introduced in refs.~\cite{Amore07a,Amore07b} to obtain a representation of the
Helmholtz equation on an uniform grid. The problem thus reduces to diagonalizing a $(N-1)^2\times (N-1)^2$ square matrix, $N-1$ 
being the number of collocation points in each direction.  We have tested the method on several examples taken from the literature. 
The application of the method is straightforward and it provides quite accurate results even for grids with 
moderate values of $N$. 
The readers interested to looking to more examples of application of this method should check the site
\begin{verbatim*}
http://fejer.ucol.mx/paolo/drum
\end{verbatim*}
where images of modes of vibration of membranes of different shapes can be found.

\end{document}